\documentclass[useAMS]{mn2e}
\usepackage{graphicx}

\title [Induced baryonic fluctuations] {Recombination induced thermodynamic Gaussian cosmological
baryonic fluctuations.}

\author[X. Hernandez] {X. Hernandez\\ 
Instituto de Astronom\'{\i}a, Universidad Nacional Aut\'{o}noma de M\'{e}xico,
Apartado Postal 70--264 C.P. 04510 M\'exico D.F. M\'exico. \\
}

\date{Released 23 October 2017}

\begin{document}

\maketitle

\begin{abstract}  
  In some instances, e.g. near phase transitions, thermodynamic fluctuations become macroscopically relevant, and relative
  amplitudes grow far above the standard $N^{-1/2}$ scale, with $N$ the number of particles. Such large fluctuations are
  characterised by a scale invariant Gaussian power spectrum. In this letter I show that the abrupt drop in the baryonic
  sound speed across recombination leads to conditions resulting in such large thermodynamic Gaussian fluctuations in the
  ionisation fraction of the baryons. Under pressure equilibrium, this will result in a mechanism for generating scale
  invariant density and temperature fluctuations in the baryonic component, inherent to the thermodynamics of the baryons
  themselves. Within a $\Lambda$CDM framework, this extra random fluctuation source leads to a decoupling of the inflationary
  relic small wave number spectrum and the amplitude of the Gaussian random fluctuations at frequencies higher than the first
  acoustic peak, an effect which could explain the mismatch between cosmic microwave background (CMB) inferences and local
  kinetic determinations of the Hubble constant. Within modified gravity theories in absence of dark matter, the mechanism
  proposed serves as a source for random Gaussian density fluctuations in the acoustic peak region.

\end{abstract}


\section{Introduction} 

The random Gaussian temperature fluctuations in the cosmic microwave background (CMB) first detected by the COBE satellite
in the early 1990s (Smoot et al. 1992), have now been extensively studied and characterised by subsequent space observatories,
in particular the WMAP (Bennett et al. 2003) and Plank missions (e.g. Plank Collaboration 2016). It is well established that
a random component having a very close to Gaussian power spectrum is present at around the recombination redshift, $z_{re}$.
Assuming an adiabatic equation of state for the baryonic component implies a corresponding density fluctuation spectrum for
the baryons. Given the tight coupling between baryons and photons prior to recombination, when the ionisation fraction was
close to unity and the effective sound speed was of $(v_{s}/c)^{2} \approx 1/3$, the corresponding pressure ensures the rapid
erasing of any baryonic fluctuations at $z>z_{re}$. Their presence at the last scattering surface is generally understood in
the context of a gravitational coupling of the baryonic plasma and an underlying and dominant dark matter component having a
Gaussian fluctuation spectrum generated during an inflationary phase, and which survives until $z_{re}$ by virtue of having
no coupling to the radiation field beyond gravitational terms.

It is interesting that thermodynamic fluctuations have Gaussian power spectra
and a scale invariant character. In some cases, e.g. near phase transitions, anomalous thermodynamic fluctuations
grow significantly, to far exceed the usual $N^{-1/2}$ scaling in the relative amplitudes of standard thermodynamic
fluctuations, to become macroscopically relevant. By considering the relevant $\Delta \phi$ potential of ionisation
fluctuations in the baryonic component near recombination, in this letter I show that a mechanism for generating
Gaussian density fluctuations appears associated to the very rapid drop in baryonic sound speed across recombination
under pressure equilibrium, inherent to purely baryonic thermodynamical processes.

Previous studies have looked at ionisation fraction fluctuations either in the context of hydrodynamical acoustic
instabilities due to coupling of the baryons to the CMB e.g. Shaviv (1998) or Liu et al. (2001) and Singh \& Ma (2002)
including a three level atom approximation, or ionisation fraction fluctuations resulting from the baryon density
fluctuations and the dependence of recombination and ionisation rates on density, Novosyadlyj (2006). In all such studies,
the resulting ionisation fluctuations are shown to be negligible.

The appearance of a further fluctuation production mechanism around recombination effectively decouples the baryonic power
spectrum in the acoustic peak region from that observed at shorter wave numbers, thought to bear the imprint and normalisation
of physics at the inflationary epoch at scales beyond the causal horizon at $z_{re}$. This extra parameter in the modeling
of CMB fluctuations might help to alleviate the present offset of over $3\sigma$ between CMB inferences and more local
expansion probes in the determination of the Hubble parameter (e.g. Riess et al. 2016, Casertano et al. 2017),
through allowing for changes in the normalisation within the acoustic peak region.

Within the context of modified gravity theories not including the dark matter hypothesis currently being explored,
e.g. MOND of Milgrom (1983), F(R) modifications such as Mendoza et al. (2013) or variants reviewed in Capozziello
\& de Laurentis (2011), the emergent gravity proposal of Verlinde (2016) or the constant bounding curvature criterion
of Hernandez et al. (2017), the requirement for a mechanism to explain the random component of observed temperature
fluctuations in the CMB becomes critical.

The precise power spectrum before the appearance of the resonant acoustic peaks will depend on the details of the
gravitational forcing term, either through a standard dark matter potential, or an enhanced baryonic modified gravity
term, and the duration of the $\Delta \phi \approx 0$ phase in units of the scale dependent sound crossing time.
Thus, although the amplitudes of the resulting density fluctuations can not be estimated easily and are absent from
this first order presentation of the phenomenon, I here show that fluctuations in the ionisation fraction, $x$, of
only $10^{-7}$ in $\Delta x$ towards the end of the recombination epoch are sufficient to explain random Gaussian
$\Delta \rho /\rho$ baryonic fluctuations of one part in $10^{5}$, at scales corresponding to the acoustic peak region
of the spectrum and smaller.

\section{Ionisation fluctuations under pressure equilibrium}

In general, thermodynamic fluctuations will occur with a probability, $\omega$, given by:

\begin{equation}
\omega=e^{-\Delta \phi/kT},  
\end{equation}

\noindent where $\Delta \phi$ is the energy associated with the fluctuation being analysed, $k$
gives the Boltzmann constant and $T$ is the temperature of the system. As it is well known, when approaching
a phase transition, the flatness of the relevant potential leads the effective $\Delta \phi$
to tend to zero, and hence the probability of the fluctuations appearing tends to unity. This yields
the observed anomalous growth of fluctuations to many orders of magnitude above the usual $N^{-1/2}$
scale on approaching critical points, e.g. the Ginsburg-Landau theory.
Although cosmological recombination is not strictly a phase transition, the extremely abrupt nature of
the process, which can be described through a $x \propto (1+z)^{12.75}$ scaling, as first estimated by Jones \& Wise (1985),
leads to the expectation of extremely flat potentials associated with fluctuations in $x$, the ionisation
fraction. Should a $\Delta \phi=0$ phase appear, we would expect the appearance of large baryonic fluctuations
near the recombination epoch, intrinsic to purely baryonic thermodynamics.

If at some point during the recombination epoch a thermodynamic fluctuation occurs such that
the unperturbed background yields a small amount of energy to a certain region resulting in
a change of the ionisation fraction within this region of $x \rightarrow x +\Delta x$, the ensuing
enhanced coupling to the photon background will result in a slightly enhanced pressure within this
region of $P \rightarrow P + \Delta P$. Imposing pressure equilibrium implies that the slightly over
ionised region will expand against the background a little, and undergo a change in its volume of
$V \rightarrow V+\Delta V$. The net energy variation of the region in question to first order in the
fluctuation will now be:

\begin{equation}
\Delta \phi = N \Delta x \epsilon_{i} - P \Delta V.
\end{equation}

\noindent In the above $\epsilon_{i}$ is the 13.6 ev of the ionisation potential of hydrogen, $P$
gives the pressure of the background and $N$ the number of atoms of hydrogen in the region in question.
We can eliminate $\Delta V$ in the above equation in favour of $M$, $\rho$ and $\Delta \rho$, the total mass,
density and density fluctuation of the region in question, through deriving $\rho =M/V$ at constant mass:

\begin{equation}
\Delta V = -\frac{M}{\rho^{2}} \Delta \rho.
\end{equation}

\noindent If we now substitute $M=N m_{p}$, approximating the total mass of the perturbed region as the number
of protons it contains, equation (2) reads:

\begin{equation}
\Delta \phi = N \Delta x \epsilon_{i} + N m_{p} P \frac{\Delta \rho}{\rho^{2}}.
\end{equation}

\noindent Notice that a positive $\Delta x$ fluctuation will result in a positive $\Delta P $ fluctuation,
expansion and hence a dilution, a negative $\Delta \rho$ fluctuation hence allowing for a $\Delta \phi =0$
point in the above relation. This is in fact the trend found in Venumadhav \& Hirata (2015), where positive $x$
fluctuations correspond to rarefactions in the density field. Such a $\Delta \phi =0$ condition will be met provided:

\begin{equation}
\frac{\epsilon_{i}}{m_{p}} = -\frac{P}{\rho^{2}}\frac{\partial \rho}{\partial x},  
\end{equation}

\noindent where I have taken the limit in the change in density of the fluctuation resulting from 
changes in the ionisation fraction only. Since $N$ has cancelled out, fluctuations of all scales will appear
if the above equation is satisfied. A symmetric situation appears for negative fluctuations, which
will appear as both enhancements and drops with respect to the mean ionisation level.
Indeed, it is a generic feature of thermodynamic fluctuations near critical points that a scale
invariant Gaussian power spectrum of fluctuations results, e.g. Landau \& Lifshitz (1980).

We can now divide both sides of the previous equation by the square of the speed of light and approximate
the adiabatic sound speed of the baryons as $v^{2}_{s}=P/\rho$ such that the $\Delta \phi =0$ condition yields:

\begin{equation}
\frac{\epsilon_{i}}{m_{p}c^{2}}= -\left( \frac{v_{s}}{c} \right)^{2} \frac{1}{\rho} \frac{\partial \rho}{\partial x}.
\end{equation}

\noindent The left hand side of the above clearly dimensionless relation has a constant value given by dividing
the 13.6 ev of the ionisation potential of hydrogen by the 938 Mev of the proton mass, which gives $1.45\times 10^{-8}$.
Whilst the left hand side of equation (6) is clearly constant, the right hand side changes dramatically across the
recombination region, where the $(v_{s}/c)^{2}$ term changes from close to $1/3$ at the beginning of recombination to
about $10^{-10}$ towards the end where $v_{s}$ settles to close to 3km/s e.g. Longair (2008).

We can now estimate $\partial \rho/ \partial x$ as usual from the background properties, by taking the total
derivative of $\rho$ as the sum of its partial derivatives, $d \rho= (\partial \rho /\partial x)dx +(\partial \rho /
\partial z)dz$ and dividing by $dx$ to obtain:

\begin{equation}
\frac{\partial \rho}{\partial x} = \frac{d \rho}{dx} -\frac{\partial \rho }{\partial z}\frac{dz}{dx}.
\end{equation}

\noindent Given the small total change in density and the large total change in $x$ over recombination, we can neglect
the first term on the right hand side of the above relation to get the $\Delta \phi =0$ condition as:

\begin{equation}
1.45\times 10^{-8} = \left( \frac{v_{s}}{c} \right)^{2} \frac{3}{z} \frac{dz}{dx},
\end{equation}

\noindent where in the above, since all this is occuring at redshift values of $z\approx 1000$, we have taken the baryon
density as $\rho=\rho_{0}(1+z)^{3} \simeq \rho_{0}z^{3}$. 

\begin{figure}
\includegraphics[width=9.0cm,height=7.0cm]{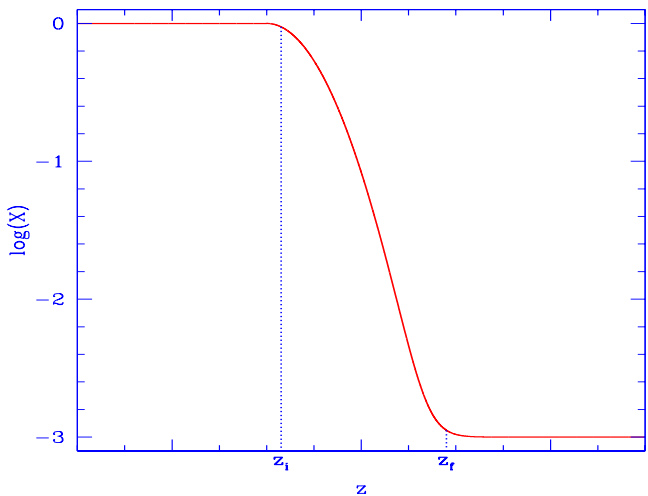}
\caption{Schematic plot of the ionisation fraction across the recombination period.}
\end{figure}

Although during the rapidly evolving phase of the recombination process $x$ scales with $(1+z)^{12.75}$ (Jones \& Wyse 1985),
we can consider a redshift, $z_{i}$, at the start of the process where the ionisation fraction begins to drop and a
behaviour close to $x=z_{3}^{n_{i}}$ will appear, with $z_{3}=z/1000$ and $0 \leq n_{i} \leq 1$, whilst towards the end
of recombination, where the ionisation fraction begins to settle to the residual value of about $10^{-3}$ a behaviour close
to $x=10^{-3}+z_{3}^{n_{f}}$ with $n_{f} \geq 1 $ will appear. A schematic representation of the above is shown in figure (1).

Thus, it is reasonable to expect that around both $z_{i}$ and $z_{f}$, $(dz/dx)$ will be of order unity. With this last
condition we can estimate the value of the right hand side of equation (8) both around $z_{i}$ and $z_{f}$.
Taking $v_{s}(z_{i}) \approx 3^{-1/2}c$,  $v_{s}(z_{f}) \approx 3 km/s$ and $z_{i} \approx z_{f} \approx 1000$
give for the right hand side of equation (8) at $z=z_{i}$ a value of $10^{-3}$, five orders of magnitude
greater than the left hand side of that relation. Similarly for $z=z_{f}$ we get a value of $3\times 10^{-13}$ for the
right hand side of equation (8), this time five orders of magnitude below the constant value of the left hand side
of this equation. Thus, although the $\Delta \phi =0$ condition is not met, by many orders of magnitude, neither
at the start nor at the end of the recombination period, the clear change from the right hand side of equation (8)
dominating near $z_{i}$ to it being the left hand side of that equation which dominates towards $z_{f}$, implies
by the intermediate value theorem, that there will necessarily be at least one point where the condition $\Delta \phi =0$
is satisfied for a critical $z_{c}$ such that $z_{i} < z_{c} < z_{f}$.

Notice also that although a number of approximations have been introduced (beyond the ones already explicitly mentioned,
including the rich atomic physics phenomenology beyond ground-level recombination, e.g. the three-level approximation of
hydrogen and helium atoms -Matsuda et al. 1971, Krolik 1990 or Hummer \& Storey 1998- or the multilevel structure considered
by e.g. Seager et al. 2000), the overwhelming difference between the left and right sides of equation (8), of five orders
of magnitude, and in opposite directions, near the start and end of recombination makes it inevitable that the condition
$\Delta \phi =0$ will be met at some point during the recombination process. At that point, anomalous fluctuations in the
ionisation fraction will appear, accompanied by corresponding density fluctuations in the baryonic density field of:

\begin{equation}
\frac{\Delta \rho}{\rho} = \frac{\epsilon_{i} \rho}{m_{p} P} \Delta x,
\end{equation}

\noindent as implied by equation (5), or within the $v^{2}_{s}=P/\rho$ approximation,

\begin{equation}
\frac{\Delta \rho}{\rho} = \frac{\epsilon_{i}}{m_{p} c^{2}} \left( \frac{c}{v_{s}}\right)^{2} x \left( \frac{\Delta x}{x}\right),
\end{equation}  

\noindent Towards $z_{f}$ where $x\approx 10^{-3}$ and $v_{s}\approx 3 km/s$, obtaining $\Delta \rho /\rho \approx 10^{-5}$
requires only $\Delta x/x \approx 10^{-4}$, i.e. $x=10^{-3} \pm 10^{-7}$, extremely small ionisation fluctuations.
Under the usual adiabatic assumption, these density fluctuations will give rise to the observed Gaussian component
in the temperature fluctuations in the CMB, of the same order.

It is clear that in the development presented the energy fluctuation giving rise to $x \rightarrow x+\Delta x$ and the
resulting volume increase of $V\rightarrow V +\Delta V$ do not occur sequentially, but adiabatically and simultaneously.
The efficiency of the mechanism proposed will hence sensitively depend on the rate at which the $\Delta \phi =0$
condition is crossed, being this efficiency maximal for a very tangential and gradual crossing which allows time for the
fluctuations described to develop at various scales. If, on the other hand, the $\Delta \phi =0$ condition is only
very briefly met, then the fluctuations described will only begin to develop at the smallest scales.

\section{Conclusions}
I have shown that in a process akin to the growth of fluctuations in the vicinity of phase
transitions, near the extremely fast phase of cosmological recombination,
the relevant potential for ionisation fraction fluctuations under pressure equilibrium will have a zero
point and hence lead to significant thermodynamic fluctuations in the baryonic density field.
As a generic feature of thermodynamic fluctuations, these will be characterised by a scale invariant
Gaussian power spectrum, as it is inferred from satellite observations of the CMB. Baryonic thermodynamical
processes during recombination lead to naturally arising random fluctuations of the type observed in the CMB. 

Within the context of a GR $\Lambda$CDM cosmology, this constitutes a further mechanism for producing
baryonic fluctuations close to the surface of last scattering. Within the context of modified gravity
theories modelling the universe without the dark matter hypothesis, the mechanism
presented here allows for an understanding of the small temperature fluctuations detected in the CMB, in
the absence of a hypothetical dominant dark matter component carrying the imprint of a remote inflationary phase
in the form of density fluctuations shielded from photon damping by the assumed lack of interaction between
dark matter and radiation.

\section*{acknowledgements}

This work was supported in part by DGAPA-UNAM PAPIIT IN-104517 and CONACyT.

\end{document}